\documentclass[iop,apj]{emulateapj}

\usepackage{apjfonts}

\newcommand{\beq}{\begin{equation}}
\newcommand{\eeq}{\end{equation}}
\newcommand{\bea}{\begin{eqnarray}}
\newcommand{\eea}{\end{eqnarray}}

\usepackage{ifpdf}
\ifpdf
\else
\fi

\begin{document}

\title{Confined Population III Enrichment and the Prospects for
  Prompt Second-Generation Star Formation}

\author{Jeremy S.\ Ritter \altaffilmark{1},
Chalence Safranek-Shrader \altaffilmark{1},
Orly Gnat \altaffilmark{2},
Milo\v s Milosavljevi\'c \altaffilmark{1},
and
Volker Bromm \altaffilmark{1} }
\altaffiltext{1}{Department of Astronomy, University of Texas, 1
  University Station C1400, Austin, TX 78712.}
\altaffiltext{2}{Racah Institute of Physics, The Hebrew University of
  Jerusalem, 91904, Israel.}

\righthead{}
\lefthead{}

\begin{abstract}

It is widely recognized that nucleosynthetic output of the first,
Population III supernovae was a catalyst defining the character of 
subsequent stellar generations. Most of the work on the earliest enrichment was
carried out assuming that the first stars were extremely massive
and that the associated supernovae were unusually energetic, enough to
completely unbind the baryons in the host cosmic minihalo and disperse
the synthesized metals into the intergalactic medium.  Recent
work, however, suggests that the first stars may in fact have
been somewhat less massive, with a characteristic mass scale of a few
tens of solar masses.  We
present a cosmological simulation following the transport
of the metals synthesized in a Population III supernova assuming that
it had an energy of $10^{51}\,\textrm{ergs}$, compatible with standard
Type II supernovae.  A young supernova remnant is inserted in the
first star's relic \ion{H}{2} region in the free
expansion phase and is followed for $40\,\textrm{Myr}$ employing 
adaptive mesh refinement and Lagrangian tracer particle
techniques.  
The supernova remnant remains partially trapped within the 
minihalo and the thin snowplow shell develops pronounced instability and
fingering.  
Roughly half of the ejecta turn around and fall back toward
the center of the halo, with $1\%$ of the ejecta reaching the center
in $\sim 30\,\textrm{kyr}$ and $10\%$ in $\sim 10\,\textrm{Myr}$.
The average metallicity of the combined returning ejecta and the
pristine filaments feeding into the halo center from the cosmic web is
$\sim 0.001-0.01\,Z_\odot$, but the two remain unmixed until accreting
onto the
central hydrostatic core that is unresolved at the end of the simulation.  We conclude that if Population
III stars had less extreme masses, they promptly enriched the host
minihalos with metals and triggered Population II star formation.  


\end{abstract}

\section{Introduction}
\label{sec:introduction}
\setcounter{footnote}{0}

The first stars forming from metal-free gas in the
early universe, the Population III (Pop III),
profoundly transformed their cosmic environment.  Radiation from these
stars first ionized the otherwise neutral, chemically pristine
gas.  In their final demise, likely involving supernovae or
the collapse into a black hole, the Pop III stars injected momentum, energy, and nucleosynthetic
products into the environment \citep{Karlsson:12}.  The amplitude and character of
these effects depended sensitively on the Pop III stellar masses, as
well as on theoretically uncertain aspects of stellar evolution. 
Early numerical investigations of the formation of Pop III stars in metal-free
cosmic halos concluded that the stars were isolated and rather massive, with masses
$\gtrsim 100\,M_\odot$
\citep{Bromm:99,Bromm:02,Abel:00,Abel:02,Gao:07,OShea:07}. Recent numerical investigations, however, have shown that angular-momentum-aided
fragmentation \citep{Stacy:10,Clark:11,Greif:11,Greif:12} and radiative termination of
accretion \citep[][see, also, \citealt{McKee:08} for an analytical approach]{Hosokawa:11,Stacy:12a} during protostellar accretion
could limit the masses to as low as
a few tens of solar masses.   The final stellar mass range remains
uncertain but could plausibly be
$\sim 1-50\,M_\odot$ with a characteristic mass $\sim 10\,M_\odot$. 

For a wide range of stellar
masses and ambient gas densities, Pop III
stars have sufficient luminosities and effective temperatures
to form extended \ion{H}{2}
regions, unless they form in especially high mass halos 
\citep[e.g.,][]{Whalen:08b} or with unusually high accretion rates
during the protostellar phase \citep[e.g.,][]{Hosokawa:12}. 
The raised pressure of the interior ionized gas produces 
a hydrodynamic response that reduces the circumstellar
gas density, and it is
the density profile modified by \ion{H}{2} region dynamics that
the supernova blastwave expands into.
The density structure imprinted by
the \ion{H}{2} region can interact
with the blastwave to drive 
turbulence and compositional mixing of the supernova ejecta.

Because the stellar ionizing photon production rate is a strong
function of stellar mass \citep[e.g.,][]{Bromm:01,Schaerer:02}, the
properties of \ion{H}{2} regions of the first stars will depend on the
mass.  A metal-free star with mass $> 100\,M_\odot$ could emit
$> 10^{50}$ ionizing photons per second, which would ionize not only
the parent cosmic minihalo, but a larger volume of
the primordial intergalactic medium (IGM).   The pressure of the
photoionized gas drives a supersonic, radially propagating pressure wave,
which ultimately shocks \citep{Whalen:04,Whalen:08b,Kitayama:04,Alvarez:06,Abel:07,Yoshida:07}.  This expels a fraction of the
baryons from the halo even before the supernova explodes.  Only
the densest clumps residing in the filaments streaming into the halo
from the cosmic web are able to avoid ionization
\citep{Abel:07}. Furthermore,
the \ion{H}{2} regions may also be prone to violent dynamical instabilities
that could create a clumpy medium \citep{Whalen:08a}, with interesting
consequences for supernova blastwave dynamics.   Most of
these features should carry over to lower mass Pop III stars with
lower ionizing photon production rates, but the
magnitude of the effects is weaker: the temperature of the
photoionized gas (at a fixed distance from the star) is lower, the pressure wave is less
pronounced, and the final ionized mass fraction inside the halo and the ionizing
photon escape fraction from the halo are lower.

Similarly, the outcome of the stellar demise is expected to depend
sensitively on stellar mass.
Restricting to nonrotating, metal-free stars with minimum mass loss
\citep[e.g.,][]{Heger:02,Heger:03}, it is expected that the stars with masses $15 - 50\,
M_\sun$ undergo core collapse, leaving behind a
neutron star, or with fallback, a black hole.  The stars
with masses $50 - 140\, M_\sun$ can produce a black hole
directly, though the stars above $100\, M_\sun$  may be affected by
pair-pulsational instability.  The stars with masses $140 - 260\,
M_\sun$ explode as pair-instability supernovae (PISNe), and even more massive
stars can also directly collapse into a black hole.  Stellar rotation
introduces new effects and significant theoretical uncertainties, e.g.,
the stars might first form a central black hole and the subsequent
accretion of the rotating stellar envelope onto the black hole might power an outflow that could
produce an explosion \citep[see,
e.g.,][]{MacFadyen:99,MacFadyen:01,Kohri:05,Milosavljevic:12,Lindner:12},
with a peculiar nucleosynthetic imprint that can be sought in especially
metal-poor stars \citep[e.g.,][]{Iwamoto:05}.

Much of the theoretical work on the formation of the
first stars and galaxies \citep[e.g.,][]{Wise:08a,Wise:12,Greif:10}, and on strategies for detecting metal-free
stellar populations in the observations of galaxies and supernovae \citep[e.g.,][]{Schaerer:03,Johnson:09,Raiter:10,Zackrisson:11}, has postulated the massive
character of metal-free stars.  
The prospect of PISNe has been particularly interesting in view of 
the large expected explosion energy $\sim 10^{53}\,\textrm{ergs}$ and
large nucleosynthetic yield, and many 
investigations have focused on this type of explosion \citep[see, e.g.,][]{Pan:12,Hummel:12,Whalen:12b}.  Such an explosion
further reduces the residual baryonic content of the halo \citep{Bromm:03b} and the metals are dispersed far
outside, thus enriching a relatively large region of the surrounding
medium \citep[e.g.,][]{Wise:08b,Greif:10,Maio:10,Wise:12}. While this
enrichment has only a minor effect on the ultimate metal budget of
the IGM \citep{Scannapieco:02,Yoshida:04}, it 
has the potential to raise the metallicity of larger collapsing halos
above a ``floor'' \citep[e.g.,][]{Bromm:03a,Smith:08}, itself a
function of the dust content \citep[e.g.,][]{Schneider:12},
 thought to be necessary for fragmentation into
low-mass stars.

There has been much less work done on 
the cosmic imprint of Pop III stars with masses in the
range tentatively favored by recent investigations \citep{Stacy:10,Stacy:12a,Clark:11,Greif:11,Hosokawa:11}.
The ionizing luminosities of these moderate-mass Pop III stars could be up to an order of magnitude lower than
those of their massive counterparts, the supernova energies
could be $\sim10^{51}\,\textrm{ergs}$, the nucleosynthetic yields could be 
at most a few solar masses per supernova, and the characteristic
abundance patterns could be more consistent with currently available
observational data
\citep[e.g.,][]{Tumlinson:06,Vangioni:11}.  Thus
the evolution of 
minihalos forming only moderate mass Pop III stars should differ in critical ways from
that of objects forming very massive ones.

Here we report the first results from
a program to study the detailed transport of the nucleosynthetic output
of the first stars, from the point of injection in supernovae, until
at least a fraction of
the new metals have recondensed in descendent cosmic
halos where they may contribute to the formation of second-generation,
metal-enriched stars.  
The specific aim of the paper  
 is to investigate the implications of the hypothesis that the first
stars had moderate masses---masses that are 
insufficient to produce extraordinarily extended \ion{H}{2} regions and ultra-energetic
supernovae. We present a cosmological simulation tracking the
supernova ejecta of one such intermediate-mass Pop III star.
It is expected on theoretical grounds that the character of the new star
formation in the descendent halo will depend on the rate of gas inflow
$\dot M_{\rm inf}$
into the nucleus of the halo, where the gas can rapidly cool and
fragment into stars.  Also, the cooling rate of the gas and the
compressibility of the turbulence in the flow depend on the average metallicity 
$Z_{\rm inf}$ and the degree of metallicity homogeneity in the
gas.  The simulation allows us to measure these parameters and thus shed light on the
conditions for the formation of metal-enriched stars in the aftermath
of a Pop III supernova.  

Our focus is the gross dynamics of the supernova remnant, and to trace the
dispersal of the metals produced. We here do not attempt to make detailed predictions for
chemical abundance patterns in second-generation stars that may in part survive to
the present day, and that can be compared with the extensive data on metal-poor stars in
the Milky Way \cite[e.g.,][]{Cayrel:04,Beers:05,Frebel:05,Lai:08}.
It has been argued that the observed abundances in these ancient stars can be explained with
Pop III supernovae from moderate-mass ($\sim 15-40\, M_{\odot}$) progenitors \citep[e.g.,][]{Joggerst:10,Tumlinson:10}. This interpretation resonates with the basic assumption on the Pop~III mass-scale made
here. In view of possibly strong selection effects in the existent metal-poor abundance data, it
is important, however, to keep an open mind in this regard. One key selection bias might result from
the high level of enrichment in the immediate neighborhood of a PISN
\citep{Wise:08b,Greif:10}, such that second-generation stars could already exhibit overall metallicities
of $10^{-3}-10^{-2}\, Z_\odot$; most surveys targeting the extremely metal-poor tail would then miss
such PISN-enriched fossils \citep{Karlsson:08}. Future surveys, such as {\it Gaia}, should provide
us with essentially unbiased samples and an opportunity to constrain theoretical models.

The paper is organized as follows.  In Section \ref{sec:simulations},
we describe our numerical algorithm.  In Section
\ref{sec:results}, we present our results.  In Section
\ref{sec:discussion}, we discuss implications for the formation of metal-enriched
star clusters in the early universe, and in Section \ref{sec:conclusions},
we present our main conclusions.  We adopt the cosmological parameters
consistent with the {\it Wilkinson Microwave Anisotropy Probe}
seven-year data \citep{Komatsu:11}.  We explicitly indicate whether
comoving or physical length units are utilized; all other quantities,
such as density, are expressed in physical units.

\section{Simulating a Supernova in a Minihalo}
\label{sec:simulations}

\subsection{Cosmological Initial Conditions and Gravity}
\label{sec:initial_conditions_gravity}

We initialize our simulation in a periodic cosmological box of volume
$1\,\textrm{Mpc}^3$ (comoving) at redshift $z=146$, large enough not only to permit the formation
of such a star, but also to allow the subsequent growth of the
star's host halo, by accretion and merging, into a larger object that
is the potential site of second-generation star formation.   The initial
conditions are generated with the multiscale cosmological initial
conditions package {\tt GRAPHIC2}
\citep{Bertschinger:01} with two levels of nested refinement on 
top of the $128^3$ base grid.  The highest refined region had an effective
resolution of $512^3$, corresponding to a dark matter particle mass of
$230\,M_\odot$.  While we will focus on the dynamics of an initially $\sim
10^6\,M_\odot$ minihalo at redshift $z\sim 20$, the region of high refinement was centered on the
overdensity that first collapses to form a more massive
$\sim 10^8\,M_\odot$ halo at a lower redshift.  The time integration was carried out
with the multipurpose astrophysical adaptive mesh refinement (AMR)
code {\tt FLASH} \citep{Fryxell:00}, version 3.3, employing the direct
multigrid Poisson gravity solver of \citet{Ricker:08}.  The AMR
refinement criteria, which we describe in Section \ref{sec:amr} below,
are set to ensure that relevant baryonic processes are sufficiently
well resolved in the simulation.  In collapsing regions where baryons
dominate the gravitational potential, the AMR resolution length quickly becomes smaller
than the dark matter particle separation.  There, we employ a
mass- and momentum-conserving quadratic-kernel-based softening
procedure developed in
\citet{SafranekShrader:12} which is applied to the dark matter
density variable during mapping onto the computational mesh, to render the dark matter particle
contribution to the gravitational potential smooth on the scale of the
local AMR grid.

\subsection{Chemistry and Cooling}
\label{sec:chemistry}

Prior to the initialization of the \ion{H}{2} region, we integrate the full
nonequilibrium chemical network for hydrogen, helium, deuterium, and
their chemical derivatives.  The chemical and radiative cooling
updates, which
include, among other processes, the cooling or heating by the cosmic microwave background (CMB),
are operator-split from the hydrodynamic update and are
subcycled within each cell.  The chemical
reaction rates are the same as we have
summarized in \citet{SafranekShrader:10}, except that the chemical and
radiative cooling update is isochoric in the present simulation.  The
network is integrated with the Bulirsch-Stoer-type semi-implicit
extrapolation method of \citet{Bader:83}.  We test
the thermodynamic evolution of the gas in
\citet{SafranekShrader:12}.  

After turning on the ionizing source,
which we describe in Section \ref{sec:ionization} below, we
continue to integrate the nonequilibrium chemical network in the
neutral gas, but in the ionized gas, we adopt a scheme in which the
gas temperature and chemical abundances relax exponentially toward a target photoionization
equilibrium.  The equilibrium gas temperature and chemical abundances
are functions of the photoionization parameter $\xi\equiv 4\pi F/n_{\rm H}$ only,
where $F$ is the radiative energy flux and $n_{\rm H}$ is the number
density of hydrogen nuclei, and are precomputed with
the same chemical network that we integrate in the neutral gas.  The
relaxation time scale is chosen to be the true, nonequilibrium thermal time scale
$t_{\rm therm}\equiv (d\ln T/dt)^{-1}$,
defined by the instantaneous local radiative cooling and
photoionization heating rates.

After inserting the supernova remnant, the
gas can contain an arbitrary admixture of metals in addition to the
primordial elements.  The radiative point source is removed, thus
discontinuing the
photoionization
equilibrium calculation. We revert to an integration of the nonequilibrium
chemical network for the primordial species and calculate the
ionization state of the metals and their contribution to the cooling
by interpolation from precomputed tables.  At temperatures above $8000\,\textrm{K}$,
the collisional ionization equilibrium state of each of the metal species,
including C, N, O, Ne, Mg, Si, S, and Fe, is calculated separately
using methods described in \citet{Gnat:07} and \citet{Gnat:12}.  Below
$8000\,\textrm{K}$ but above $200\,\textrm{K}$, we ignore the metallic
contribution to the free electron abundance 
and tabulate a single density-independent cooling function (such that
the volumetric radiative loss rate scales with the square of the density), appropriate for a gas at
relatively low densities $\lesssim 10^3\,\textrm{cm}^{-3}$, using the
code {\tt CLOUDY} \citep{Ferland:98}.  The tabulated data extend to a
minimum metallicity of $10^{-3}\,Z_\odot$; below this metallicity, we
set the metal cooling rate in this range of temperatures to zero.  
Below $200\,\textrm{K}$, we
disable cooling by metal lines altogether; however, because
low-temperature thermodynamics is not the focus of this work, this has
no effect on the forthcoming results.  The treatment of primordial species
takes into account the electrons provided by the metals and vice
versa, but we do not include collisional charge exchange reactions between the
primordial species and the metal species.  We assume that the metals have solar abundance
ratios. This is an arbitrary, unnatural choice given that our metals
are being produced by a core-collapse supernova, but this choice
should have only a very minor effect on the thermodynamic evolution of
the supernova remnant.

\begin{figure}
\begin{center}
\includegraphics[width=0.45\textwidth]{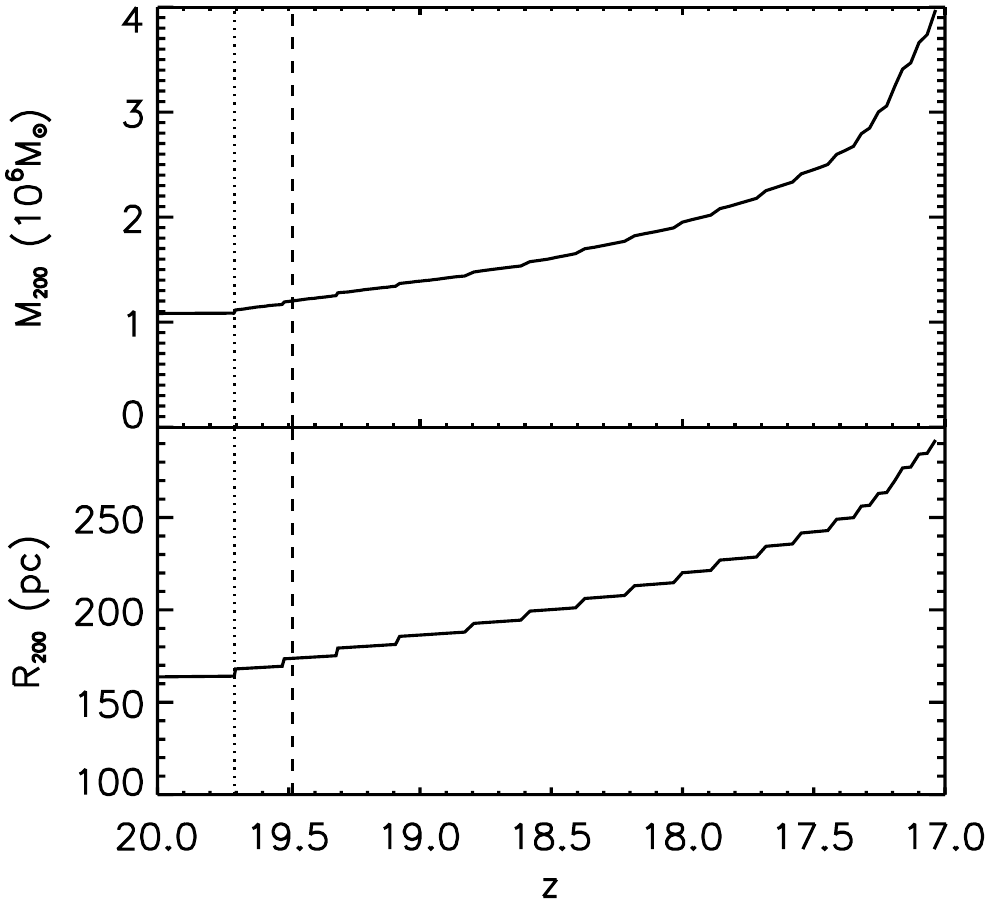}
\end{center}
\caption{The halo mass $M_{200}$ and radius $R_{200}$ as a function of
redshift, over the course of $50\,\textrm{Myr}$.  The dotted line marks the formation of the star, and dashed
line the supernova explosion. The steps are an artifact of the method
we use to estimate $R_{200}$.\label{fig:200}}
\end{figure}

\begin{figure*}
\begin{center}
\includegraphics[width=0.33\textwidth]{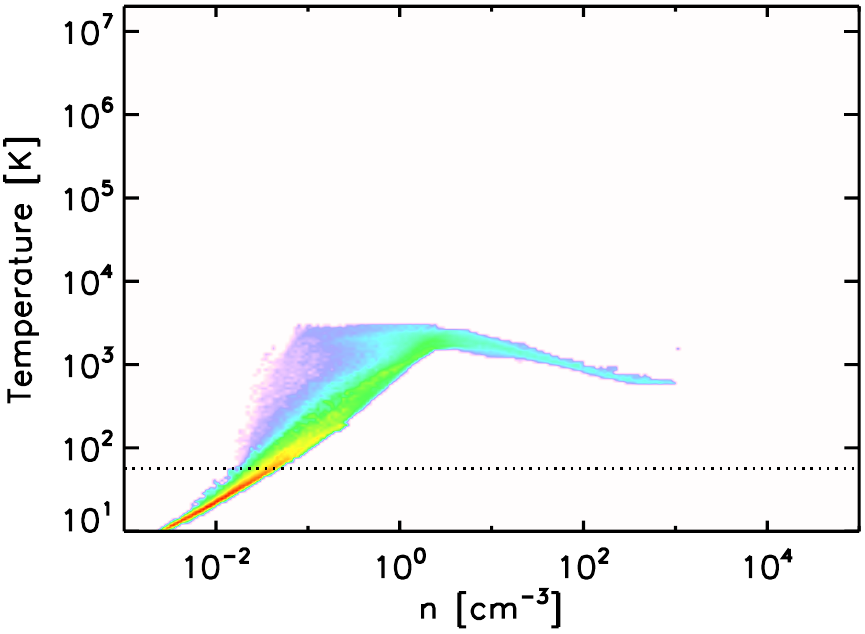}
\includegraphics[width=0.33\textwidth]{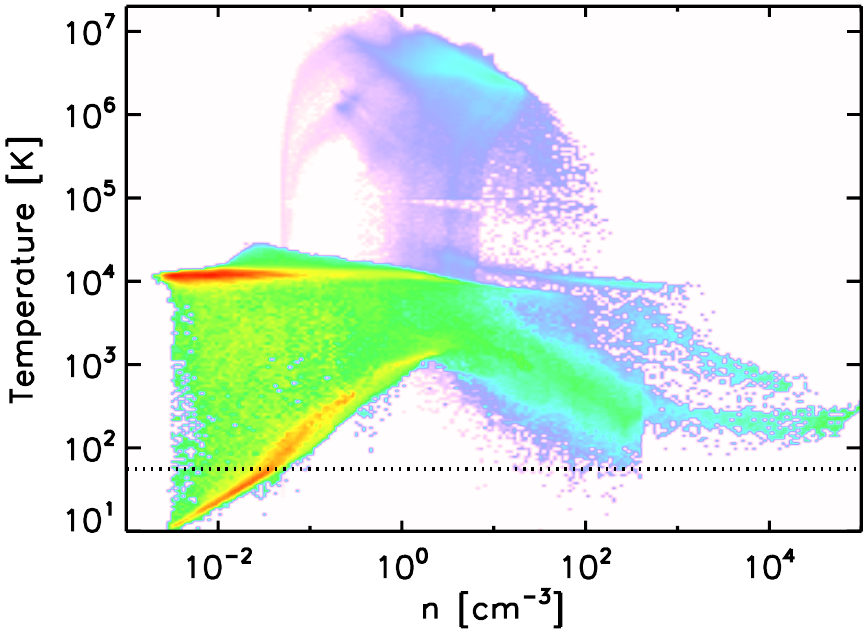}
\includegraphics[width=0.33\textwidth]{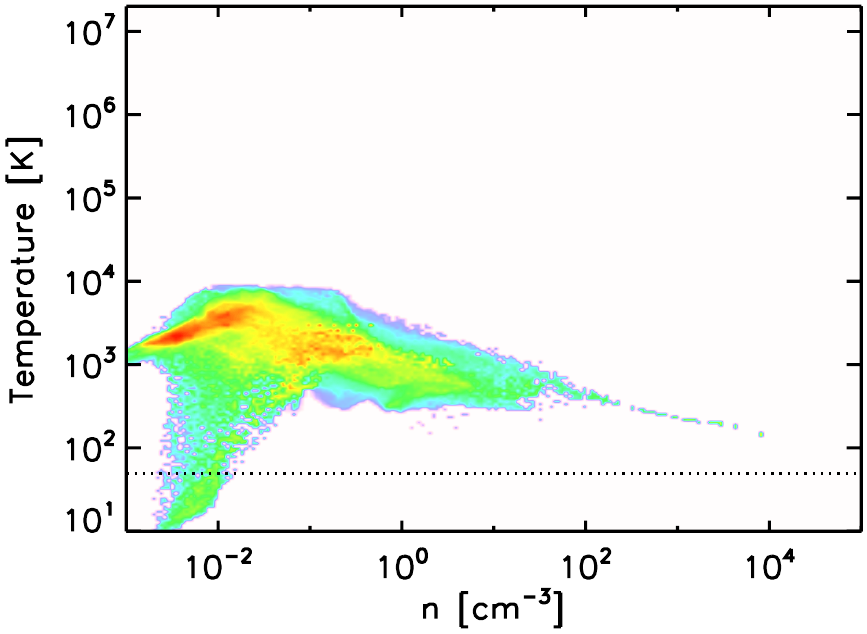}
\end{center}
\caption{Density-temperature phase plot for the gas within
  $1\,\textrm{kpc}$ (physical) from the center of the minihalo at
  redshifts, from left to right, $z=19.7$ (right before the insertion
  of the star), $z=19.5$ (in the Sedov-Taylor phase of the supernova,
  $8\,\textrm{kyr}$ after the explosion),
  and at $z=17$ (end of the simulation, $40\,\textrm{Myr}$ after the explosion).  By the end of the
  simulation, the gas heated by photoionization and the supernova has
  cooled and the halo has resumed central gravitational collapse,
albeit at four times higher halo mass and now a nonzero
metallicity. The dashed line shows the CMB temperature.\label{fig:denstemp}}
\end{figure*}

\subsection{Adaptive Mesh Refinement Strategies}
\label{sec:amr}

The simulation is set to adjust the {\tt FLASH} refinement level $\ell$ based on the local baryon
density $\rho_{\rm b}$ to ensure that denser regions are more highly
refined.  
The grid separation (cell size) is related to the
refinement level via $\Delta x = 2^{-\ell+1-3} L$, where $L$ is the
size of the computational box.\footnote{The factor of $2^{-3}$ arises from
the subdivision in {\tt FLASH} of each computational block relative to which the
refinement level is defined into $8^3$ cells.}   As
gravitational collapse drives the local fluid density $\rho_{\rm b}$
to increasingly higher values, the local refinement level is
adjusted to satisfy $\rho_{\rm b}< 3\bar \rho_{\rm b} \,
2^{3(1+\phi)(\ell-\ell_{\rm base})}$ at all times, where $\ell_{\rm
  base}=5$ is the refinement level corresponding to the base grid of size
$128=2^{\ell_{\rm base}+2}$, 
we choose the scaling parameter to equal $\phi=-0.3$ (value consistent
with \citealt{SafranekShrader:12}).  We insert a
star particle when the hydrogen number density first exceeds
$1000\,\textrm{cm}^{-3}$; this happens at redshift $z\approx19.7$ and refinement level $\ell=14$,
corresponding to cell size $\Delta x\approx 0.74\,\textrm{pc}$ (physical). 

At the end of the star's life and just prior to inserting the young
supernova remnant, we force the simulation to refine further to level
$\ell=21$ at the location of the star particle, corresponding to cell size
$\Delta x\approx 0.006\,\textrm{pc}$ (physical). This allows us to
insert the supernova remnant well within the free-expansion phase,
before the ejecta have started to interact with the circumstellar
environment.  Then, we disable automatic derefinement, and allow the
simulation to refine additional computational cells, up to the maximum
level $\ell_{\rm max}$ initially equal to $21$, based on the standard
second derivative test in {\tt FLASH}, but in this case, the second
derivative test is computed using the cell metallicity.  This
ensures that the entire region enriched with metals will be resolved at the
refinement level $\ell_{\rm max}$.  As the supernova remnant expands,
we gradually and cautiously lower $\ell_{\rm max}$, and thus coarsen the resolution
of the metal-enriched region, to allow us to use ever longer time
steps and integrate the simulation for $40\,\textrm{Myr}$
past the insertion of the supernova.

\subsection{The First Star and Ionizing Radiation Transfer}
\label{sec:ionization}

When the hydrogen number density first exceeds $1000\,\textrm{cm}^{-3}$ near
the center of the collapsing minihalo, we insert a collisionless star 
particle.\footnote{This threshold density
  for star formation is ad hoc, but consistent with the choice in,
  e.g., \citet{Wise:12}.}  
The star is free to move in the combined gravitational
potential of the dark matter and baryons but does not contribute to the
potential, i.e., it is treated as massless.  We calculate the transfer of the star's
ionizing radiation and the response of the circumstellar medium to the
associated photoheating in the \ion{H}{2} region as follows.
We assume that it is an isotropic
source of hydrogen-ionizing radiation and treat the radiation as monochromatic
with photon energy $h\nu=16\,\textrm{eV}$ and photon emission rate
$Q({\rm H})=6\times10^{49}\,\textrm{photons}\,\textrm{s}^{-1}$. This ionizing
photon rate
is about three times as high as has been estimated for a
$40\,M_\sun$ metal-free star without mass loss, $Q({\rm
  H},40\,M_\odot)\approx
2\times10^{49}\,\textrm{photons}\,\textrm{s}^{-1}$
\citep{Schaerer:02}. 
We deliberately overestimate the ionizing luminosity to allow for the
presence of additional, lower-mass, longer-lived stars also producing additional
ionizing photons in a compact cluster around the primary $40\,M_\odot$
star.  Also, boosting the ionizing luminosity allows the \ion{H}{2}
region to break out more easily at the moderate spatial resolution at which
we insert the star particle and perform ionizing radiation transfer.

Our intention was to calibrate
the monochromatic
 ionizing photon energy to
a $40\,M_\odot$ star, but the prescription we used to arrive at the adopted value
of $16\,\textrm{eV}$ suffers from lack of physical consistency.  In
retrospect, the inconsistency
could have been avoided by carrying out 
photon-number-flux-weighted averaging
of photoionization cross sections and heating rates as
in, e.g., Appendices A.2 and A.3 of \citet{Pawlik:12}, assuming a
spectral energy distribution containing a
single $40\,M_\odot$ Pop III star with an effective temperature
$\sim 8\times 10^4\,\textrm{K}$ \citep{Schaerer:02}, as well as perhaps several lower-mass 
companion Pop III stars with lower
effective temperatures. This would have resulted in higher average
photoelectron energies than in our monochromatic treatment.
The unintended consequence
of our crude prescription is that in the simulation, photoionization will
heat the gas only to a maximum temperature $\sim14,000\,\textrm{K}$,
whereas for an effective photospheric temperature of, e.g., $\sim
10^5\,\textrm{K}$ (similar in moderate mass and very massive metal
free stars), the gas at small
distances $\lesssim 1-10\,\textrm{pc}$ from the star should be heated
to higher temperatures $\sim 30,000\,\textrm{K}$, especially as the
ionized gas density drops \citep[see,
e.g.,][]{Yoshida:07}.  An artifact of this is a potential underestimate of the
amplitude of the outward-propagating pressure wave resulting from
the intense photoionization heating in the \ion{H}{2}.  We address
this issue further in Section \ref{sec:HII_region} below. 

To carry out ionizing radiation transfer, we center a spherical coordinate
system on the current position of the star particle, and pixelize the angular
coordinate using the {\tt HEALPix} scheme of \citet{Gorski:05} with $N_{\rm
  pix}=3072$ pixels.  Within each angular pixel, we split the radial
coordinate in the range $4\times10^{20}\,\textrm{cm}\leq r < 10^{24}\,\textrm{cm}$
(comoving), corresponding to $0.7\,\textrm{pc}\lesssim r
<17\,\textrm{kpc}$ (physical), into $N_{\rm rad}=103$ logarithmic radial bins.  For
each volume element $E_{p,b}$ defined by an angular {\tt HEALPix}
pixel $p$ and
a radial bin $b$, and for each mesh cell $E_c$ defined by the
cell index $c$, we compute the volume of the intersection of the two
elements $V_{p,b;c}\equiv {\rm vol}(E_{p,b} \cap E_c)\geq 0$.  In general,
only a fraction of the cell lies within $E_{p,b}$.  In this case,
we recursively split the cell into sub-cells via an 
octree subdivision procedure and compute the contribution of the subcells lying
within $E_{p,b}$ to the intersection volume; the recursion is continued until a
desired accuracy is achieved.  Armed with the intersection volumes, we
compute the average case-$B$ hydrogen recombination rate $\dot
n_{{\rm rec},p,b}$ evaluated under the assumption of full ionization (as in the
standard Str\"omgren calculation)
within each $E_{p,b}$ via
\beq
\label{eq:n_dot}
\dot n_{{\rm rec},p,b} = \frac{1}{V_{p,b}}\sum_{c} \alpha_{{\rm
    rec},B}(T_c) n_c^2 V_{p,b;c} ,
\eeq
where $V_{p,b}={\rm vol}(E_{p,b})\approx  \sum_c V_{p,b;c}$.  In
Equation (\ref{eq:n_dot}), $T_c$ and
$n_c$, are, respectively the temperature and density of hydrogen
nuclei within the cell, and $\alpha_{{\rm
    rec},B}(T)$
is the hydrogen
recombination coefficient.   Within each bin, we compute the
approximate integral of $4\pi r^2 \dot n_{{\rm rec},p}(r)$ along the radial
direction (now replacing the bin index $b$ with the continuous radial
coordinate $r$) and determine the pixel-specific Str\"omgren radius
$R_{{\rm S},p}$ that solves the equation 
\beq
Q ({\rm H}) = \int_0^{R_{{\rm S},p}} 4\pi r^2 \dot n_{{\rm rec},p}(r)
dr .
\eeq
We assume that no ionizing radiation penetrates to radii $r\geq
R_{{\rm S},p}$.  At smaller radii, we compute the local ionizing
photon number flux (in the units of
$\textrm{photons}\,\textrm{cm}^{-2}\,\textrm{s}^{-1}$) using
\beq
f_r(r) = \frac{1}{4\pi r^2} \left[Q ({\rm H})  -  \int_0^{r}
4\pi r^2 \dot n_{{\rm rec},p}(r') dr' \right] .
\eeq
The local ionizing flux is then used to compute the local
photoionization parameter for cell $c$ lying within the pixel $p$ with
a center at a radius $r_c$ via $\xi_p(r)=4\pi h\nu f_p (r_c) /n_c$.  Given a
value of the
photoionization parameter, the photoionization
equilibrium temperature of the gas and the chemical abundances within
the cell are evaluated via table lookup (see Section
\ref{sec:chemistry}).  The computation is fully parallel and is
repeated at every hydrodynamic time step.  We do not compute the radiation pressure from Thomson
scattering and photoionizations.  This radiative transfer scheme is not
explicitly photon conserving and is not designed to correctly
reproduce the ionization
front kinematic especially when an R-type ionization front is
expected, 
but it is sufficient to reproduce the basic hydrodynamic response of the
gas to photoionization heating.

\subsection{Supernova and Metal Transport}
\label{sec:supernova}

We insert the supernova $3\,\textrm{Myr}$ after inserting the star.
The supernova is initialized in the free-expansion phase, when the remnant
is about $t-t_{\rm SN}=35\,\textrm{yr}$ old, well 
before the freely expanding ejecta have
started to interact with the primordial circumstellar medium. This
early insertion is possible thanks
to our ability to drastically increase the local mesh resolution at a specific
time in the simulation---just prior to supernova explosion---and then
gradually degrade the resolution as the remnant expands. The ejecta
are initially
cold and have a total mass $M_{\rm SN}=40\,M_\odot$ and a total kinetic energy of
$E_{\rm SN}=10^{51}\,\textrm{ergs}$.  
The metal yield is assumed to be $Z_{\rm SN}=0.1$ and uniformly premixed in the
ejecta, implying $4\,M_\odot$ of metals. The
velocity of the ejecta is linear in the distance from the center of
the explosion and the density is uniform.  The initial radius of the
remnant is chosen such that its radius is $10\%$ of the radius of the
remnant at which the swept up mass equals the ejecta mass, or in this case,
$R_{\rm SN}=0.075\,\textrm{pc}$.  The simulation carries out advection
of the absolute metallicity $Z$ by treating it as a passive mass scalar
quantity.  This advection is subject to undesirable numerical
diffusion, which is unavoidable in Eulerian solvers of this type
\citep[see, e.g.,][]{Plewa:99}.  Therefore, in parallel with passive
scalar advection, we 
also insert a number $N_{\rm part}=5\times10^5$ of passive
Lagrangian tracer particles, originally uniformly distributed in the
ejecta. The simulation advances particle positions by
integrating the first-order ordinary differential equation $d{\mathbf
  X}_i/dt={\bf v}({\mathbf X}_i)$, where ${\mathbf X}_i$ is the location
of particle $i$, and ${\mathbf v}({\mathbf X}_i)$ is the baryon fluid
velocity, quadratically interpolated from the computational grid, at the location of
the particle.  We do not take into account the possibility that
the supernova may be a source of molecules and dust.  Both are undoubtedly produced in
the free-expansion phase, but their survival of the reverse shock
hinges on small-scale clumpiness of the ejecta and on other complex, incompletely-understood
physics \citep[see, e.g.,][]{Cherchneff:09,Cherchneff:10}.

\begin{figure}
\begin{center}
\includegraphics[width=0.45\textwidth]{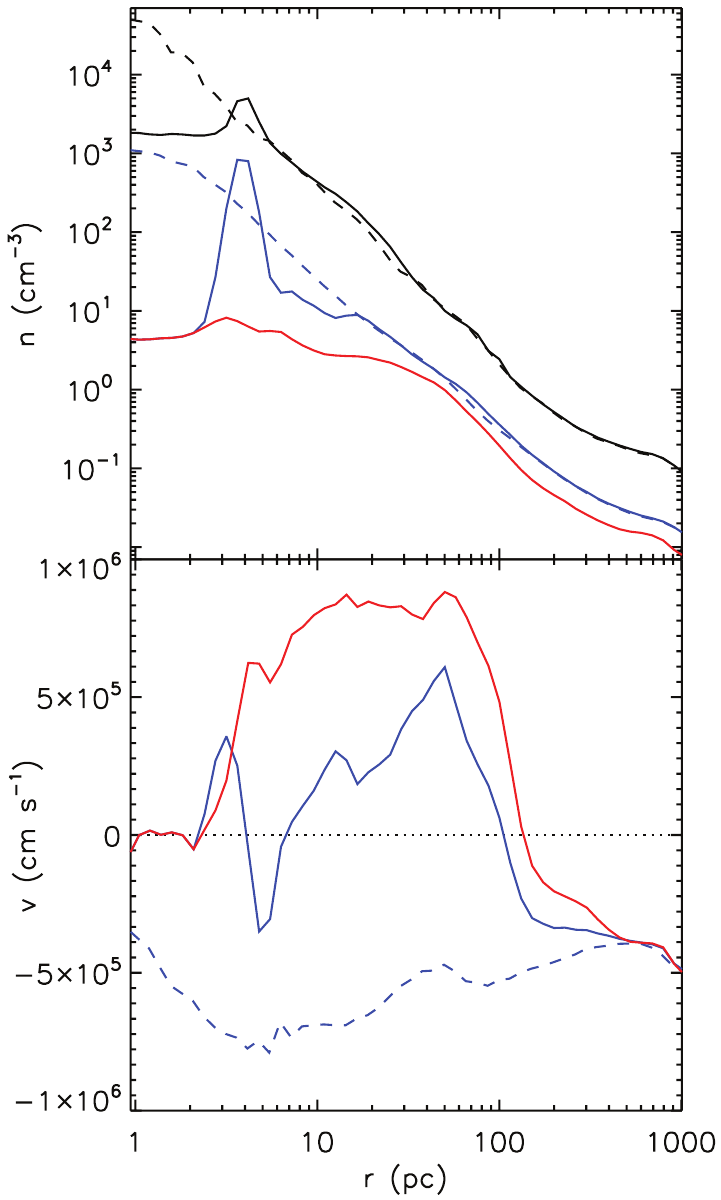}
\end{center}
\caption{Spherically averaged density (upper panel) and radial velocity
  (lower panel) profiles for all baryons (blue line) and the ionized
  gas (red line), before the insertion of the star at $z=19.7$
  (dashed lines), and before the insertion of the supernova at $z=19.5$
  (solid lines). For reference, we also show the density profile of
  gravitational mass density (including dark matter) in the units of
  the proton mass (black line).  The neutral gas density peak located
  inside the \ion{H}{2} region is a clump of dense neutral gas that
  has withstood photoionization. The gas with positive (negative)
  radial velocity is flowing away from (toward) the star.\label{fig:HII_region_profile}}
\end{figure}

\begin{figure*}
\begin{center}
\includegraphics[width=0.9\textwidth]{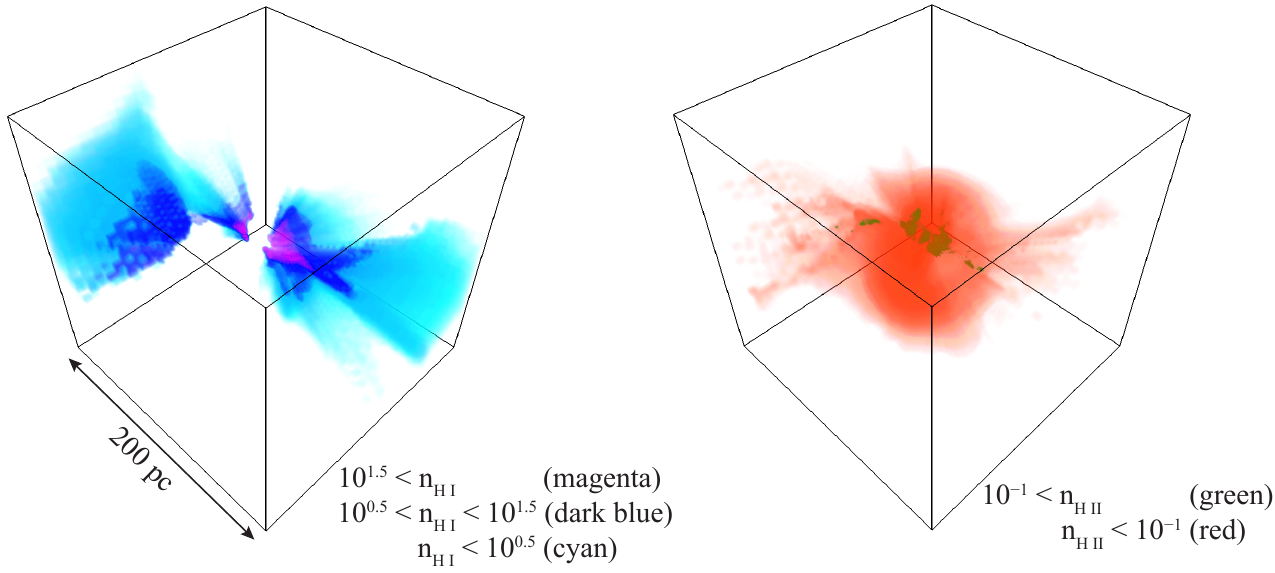}
\end{center}
\caption{Volume rendering of the number density of neutral (left
  panel) and
  ionized (right panel) hydrogen at $z=19.5$, right before the insertion
  of the supernova, in a $200\,\textrm{pc}$ cube centered on the star
  particle.  The colors show gas density ranges (in units of
  $\textrm{cm}^{-3}$) as indicated in
  the legend.  The densest ionized gas clumps $n_{\rm
    H\,II}>10^{-1}\,\textrm{cm}^{-3}$ in the center of the cube seem to be fed by
  photo-ablation from even denser, $n_{\rm
    H\,I}>10^{1.5}\,\textrm{cm}^{-3}$, persistent cold neutral clumps
  in this region.\label{fig:vapor}}
\end{figure*}

\section{Results}
\label{sec:results}

In Figure \ref{fig:200}, we show the redshift evolution of the radius
$R_{200}$ within which the average halo density equals $200$ times the
critical density, as well as the corresponding enclosed halo mass
$M_{200}$.  The star particle is inserted at $z\approx19.7$ and the
supernova at $z\approx 19.5$.  The halo radius doubles from $R_{\rm
  200}\approx 160\,\textrm{pc}$ to $300\,\textrm{pc}$ in the course of the
simulation, and the halo mass quadruples from $M_{\rm 200}\approx
10^6\,M_\odot$ at star insertion to $4\times10^6\,M_\odot$ at the end
of the simulation.  
Figure \ref{fig:denstemp} shows the density-temperature phase diagram
for gas within $1\,\textrm{kpc}$ (physical) from the center of the
halo, immediately prior to the formation of the star particle, after
the supernova explosion, and finally at the end of the simulation when
the halo mass has quadrupled.  Before stellar insertion, the
thermodynamic structure of the halo 
is that of an archetypical minihalo in which gas cools by rovibrational
transitions of molecular hydrogen.
 The thermodynamic structure at the end of the simulation is similar, which illustrates that the halo relatively
quickly ``recovers'' from the radiative and kinetic impact of the first star and
resumes central gravitational collapse.  
Photoionization and
the supernova blastwave, which we discuss in detail in Sections
\ref{sec:HII_region} and \ref{sec:blastwave}, respectively, only
temporarily disrupt the cooling and condensation of gas in the
halo center.  Then, in Section 
\ref{sec:enrichment}, we study the transport of
the metals injected by the explosion.

\subsection{The \ion{H}{2} Region}
\label{sec:HII_region}

 The photoionization raises
the temperature of the interior of the \ion{H}{2} region to $\approx
10^4\,\textrm{K}$ with a maximum temperature $\approx
1.4\times10^4\,\textrm{K}$; the latter temperature ceiling is an
artifact of our photoionization prescription (see Section
\ref{sec:ionization}).  The increase in gas pressure drives an outflow
in the ionized gas, thus lowering the central density.  
  In Figure
\ref{fig:HII_region_profile}, we show the star-centered, 
spherically-averaged gas density and spherically-averaged,
mass-weighted radial velocity profile just before
inserting the star, and $3\,\textrm{Myr}$ later,
just prior to inserting the supernova. During this interval, the
spatial resolution in the vicinity of the star is $\Delta x\approx
0.75\,\textrm{pc}$. We separately examine the
density and velocity profiles of all gas and of the ionized gas only.
The ionized gas has a maximum outward velocity of
$\sim 9\,\textrm{km}\,\textrm{s}^{-1}$, which can be compared to the
maximum gas inflow velocity $\sim - 7\,\textrm{km}\,\textrm{s}^{-1}$
before photoionization.
The central density drops from $\sim 2000\,\textrm{cm}^{-3}$ to
$\sim 4\,\textrm{cm}^{-3}$.  Interestingly, at radii
as small as $2-3\,\textrm{pc}$, the densest gas clumps ($n\sim
10-1000\,\textrm{cm}^{-3}$) resist photoionization and do not acquire
the outward velocity of the ionized gas.  

An examination of the
three-dimensional geometry of the ionized and neutral phases in Figure
\ref{fig:vapor} shows that the sustained neutral (i.e., self-shielding) clumps are associated
with the photoablated terminus of dense gas filaments feeding into the halo from
the cosmic web.  This outcome was already observed by \citet{Abel:07}
in a similar simulation of a primordial \ion{H}{2} region with
more accurate radiation transfer.  
It is worth stressing that our simulation may not properly resolve the
 process of the ionization of dense cloud
  cores because the method is not explicitly photon conserving, and
  because it may
  not resolve the shock that is expected to precede the
  D-type ionization front in the cold gas (we find that the
  ionized gas near the D-type front 
is slightly overpressured relative to the cold gas).
For these reasons, 
  we are not certain that the simulation has converged in terms of the
  rate of photoablation of dense neutral clumps, 
but qualitative consistency with previous work is reassuring.
Continuous photoablation
presents a source of fresh, dense ionized gas in the vicinity of the
star.  This, in addition to the temperature ceiling mentioned above,
explains why the central ionized gas density in our
simulation remains higher than the value $\sim
0.1\,\textrm{cm}^{-3}$ characteristically observed in one dimensional
models.  

We do not observe a clear
indication of the expected outward propagating pressure wave and the
associated shock
wave, which would have manifested as a dense, supersonically
propagating shell.  Such a shell could become important at later times as it
would modify the dynamics of the supernova blastwave; e.g., it
could drive a reverse shock into the supernova ejecta which could trigger
hydrodynamical instability and mixing of the ejecta with the swept up
circumstellar medium \citep{Whalen:08b}.
The absence of the pressure wave can be attributed to our 
approximate treatment of ionizing radiation transfer where the ionization
front speed is not properly constrained by the finite ionizing photon
consumption rate and is resolution dependent, and also because of the
artificial temperature ceiling.  However the ionized gas does acquire
a positive radial velocity out to radii $\sim 120\,\textrm{pc}$,
consistent with expectations for our adopted source luminosity.

\begin{figure*}
\begin{center}
\includegraphics[width=\textwidth]{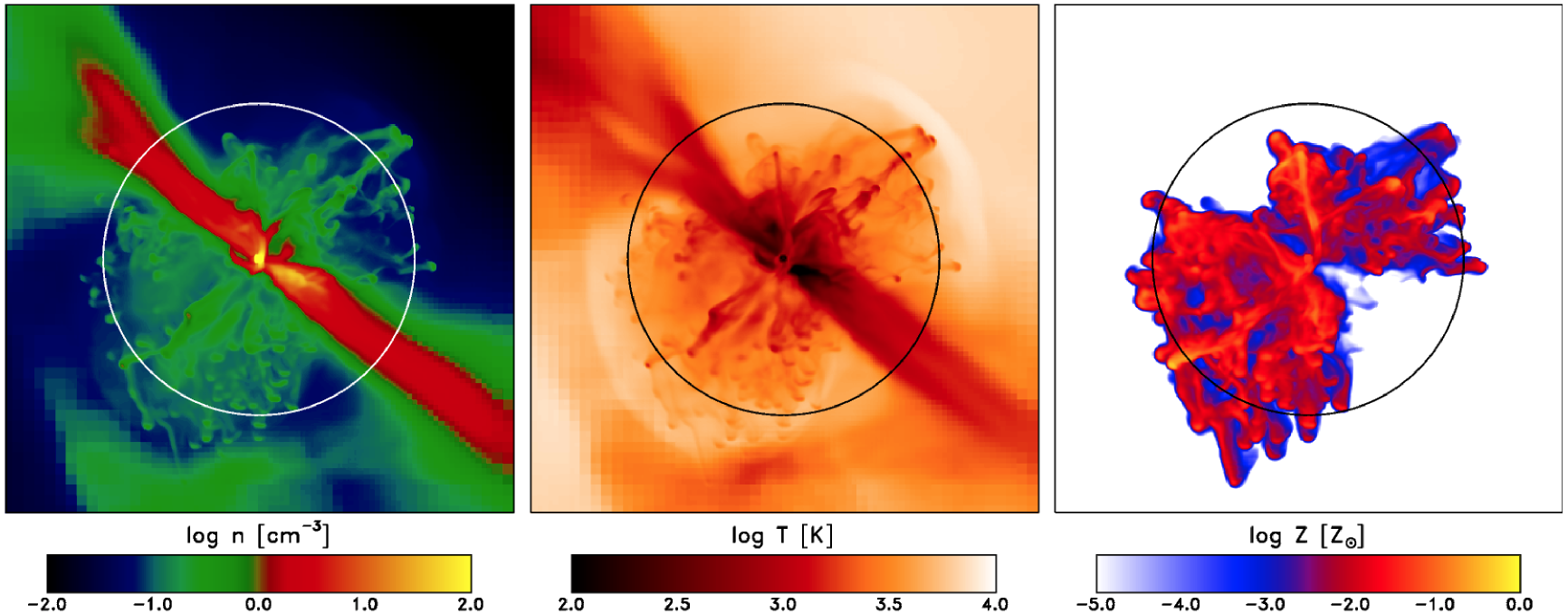}
\end{center}
\caption{Baryon mass-weighted projection of hydrogen density,
  temperature, and fluid-based metallicity in units of the solar
  metallicity $8.5\,\textrm{Myr}$ after the
  explosion.  The circle shows the virial radius of the halo, which is here
  $180\,\textrm{pc}$ (physical).\label{fig:projection}}
\end{figure*}

\subsection{Blastwave Evolution and Fallback}
\label{sec:blastwave}

Because of the low energy and compactness of the remnant, the cooling by inverse
Compton scattering off the CMB is relatively unimportant, different from the
case of ultra-energetic Pop III supernovae, where this cooling channel
dominates the early evolution \citep{Bromm:03b}. 
The blastwave also largely avoids the dense neutral clumps resisting
photoionization, 
and they remain a repository of low-entropy gas in the halo.  
This is easily seen in Figure \ref{fig:projection}, 
which shows mass-weighted
projected density, temperature, and metallicity $8.5\,\textrm{Myr}$
after the explosion.
Outside of the filaments, 
after $40\,\textrm{Myr}$, the effects of photoionization and
the blastwave have largely been erased; the thermal phase structure
within the halo is similar to that of a higher-mass unperturbed
minihalo. The appearance of warmer
($\lesssim 10^4\,\textrm{K}$) gas at low densities reflects the
higher virial mass of the halo at the end of the simulation and
the residual entropy left by the fossil \ion{H}{2} region.
The presence of metals at local metallicities $Z\gtrsim
10^{-5}\sim 10^{-3}\,Z_\odot$ and an enhanced abundance of hydrogen-deuteride 
now facilitate more rapid cooling toward low temperatures.
The enhanced cooling is now especially apparent at high densities
$n_{\rm H}\gtrsim 100\,\textrm{cm}^{-3}$ that are found in the central
region. The dense gas is insufficiently well resolved at the end of the
simulation; the resolution limit hinders compression of the fluid to
densities $n_{\rm H}\gtrsim 1000\,\textrm{cm}^{-3}$.  
Nevertheless, it is clear that the dense gas resides at temperatures a
factor of a few lower than the gas at the same densities prior to the
insertion of the ionizing source.  We do not observe the anticipated 
gas cooling to the CMB temperature, either through metal line cooling
or the cooling by hydrogen-deuteride \citep[e.g.,][]{Johnson:06},
because we enforced an artificial
temperature floor at $200\,\textrm{K}$.\footnote{The CMB temperature
  would have been a more natural choice for the temperature floor, but
at densities at which radiative cooling below our assumed temperature
floor would occur, the simulation would not be resolving the local
Jeans length and would thus in any case not be able to accurately
reproduce fragmentation in the gas.  Thus, we defer a study of the
fragmentation to a follow-up simulation.}

\begin{figure*}
\begin{center}
\includegraphics[width=0.95\textwidth]{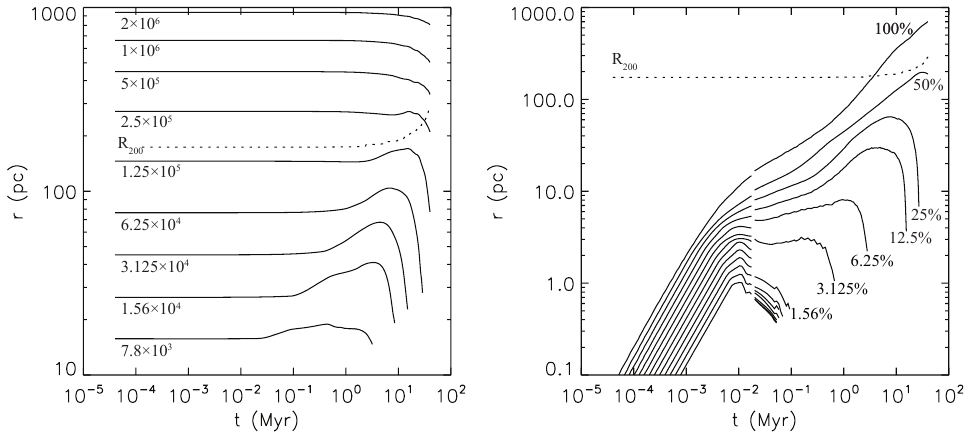}
\end{center}
\caption{Characteristic radii corresponding to Lagrangian mass coordinates
 enclosing baryonic masses of, from top to bottom,
 $(2,\,1.0,\,0.5,\,\cdot\cdot\cdot)\times10^6\,M_\odot$ (left
 panel; the masses are labeled on the curves in the units of $M_\odot$) and ejecta mass fractions, as tracked by passive tracer particles, of $(100\%,\,50\%,\,25\%,\,\cdot\cdot\cdot)$ 
(right panel; percentages are labeled on the curves), both
 as a function of time
since the explosion.  The dotted line
shows the virial radius $R_{200}$ of
the host halo.  The baryonic radii are computed from the gravitational
potential minimum of the halo.  Supernova-driven expansion is evident
in the mass coordinates inside the virial radius, but all mass
coordinates return to infall owing in part to a persistent
filamentary streaming
of fresh, cold baryons from the cosmic web into the halo center.  For
ejecta radii, at times
$t<0.02\,\textrm{Myr}$, the location
of the progenitor star is taken as the coordinate
origin; at later times, the origin
is the location of the gravitational
potential minimum; the change of origin is visible as a discontinuity
in the curves.  By the end of the simulation, a half of the ejecta
has turned around and is falling back toward the center of the
halo. At late times, the curves plunge steeply as the material
approaches the central unresolved fluid elements with a finite free-fall velocity. \label{fig:lagrangian_radii}}
\end{figure*}

An insight into the rate of recovery of the halo can be obtained from
Figure \ref{fig:lagrangian_radii}, which shows the Lagrangian
radii that track the radial
location of a grid of mass coordinates 
of the spherically averaged baryon and
supernova ejecta mass as a function of time from the explosion. 
In this period, we gradually
coarsen the spatial resolution from $\Delta x\approx
0.006\,\textrm{pc}$, a high value achieved by telescopic adaptive mesh refinement
at the location of the explosion, to $\Delta x\approx 14\,\textrm{pc}$
at the end of the simulation, $40\,\textrm{Myr}$ later.
 It is worth keeping in mind that the underlying flow is complex and contains simultaneous
inflows and outflows that are averaged over; each of the radii
reflects only the net
displacement of the mass coordinate.  The
baryonic Lagrangian radii show expansion, driven by the supernova
blastwave, starting at $\approx 20\,\textrm{kyr}$ at the innermost
radius shown, and appearing at progressively larger mass coordinates
until outward-directed motion becomes evident near the virial radius at $\approx 2\,\textrm{Myr}$ after
the explosion.  Each of the radii corresponding to mass coordinates inside
the halo, however, shows a reversal from a net outflow to a net inflow.
The reversal commences at $\approx 400\,\textrm{kyr}$ at the innermost
radius shown, and reaches progressively larger radii until $\approx
20\,\textrm{Myr}$ when the flow at the virial radius
turns into an inflow.  Outside of the virial radius, cosmic infall
persists largely unabated throughout the remnant's history.

\begin{figure*}
\begin{center}
\includegraphics[width=\textwidth]{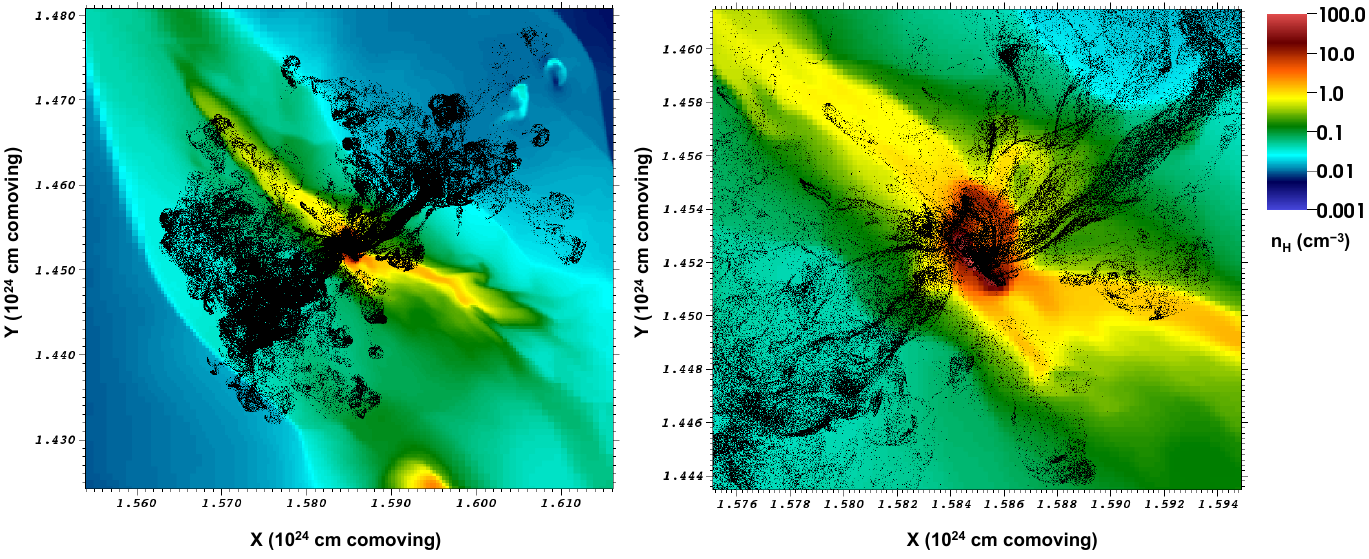}
\end{center}
\caption{Metal dispersal and fallback at the end of the simulation,
  $40\,\textrm{Myr}$ after the supernova explosion, at $z=17$. The left panel is
  $1.1\,\textrm{kpc}$ (physical) wide and centered on the gravitational
  potential minimum; the right panel is a $360\,\textrm{pc}$ detail.
  The black points are the metal particles in projection, while the
  color shows a slice through the physical hydrogen density $n_{\rm H}$.
  Metal-bearing Rayleigh-Taylor fingers have a positive radial
  velocity and have breached the virial radius of the halo, while a
  majority of the metal mass is falling back into the halo center and
  remains incompletely mixed with the primordial gas. Density in the
  central unresolved core exceeds coverage of the color scale and is $\sim10^4\,\textrm{cm}^{-3}$.\label{fig:metal_fallback}}
\end{figure*}

An
examination of the evolution of the actual three-dimensional structure
of the baryonic flow (see the projection in Figure \ref{fig:projection})
reveals that the prompt reversal can only in part be
attributed to a fallback of the mass that the blastwave has swept up
and to which it has
transferred momentum.   Instead, the filamentary inflow of dense, cold,
molecular gas from the nearby cosmic web directly into the center of the
minihalo contributes a large fraction of the mass inflow rate. The baryonic
filaments are seen diagonally from top left to bottom right in Figure
\ref{fig:metal_fallback}, which shows a slice of the density field.
Because the filaments survive photoionization and the explosion
largely intact, they can sustain a net inflow into the center of the
halo even when a majority of the ejecta and the swept up mass are
moving outward.

The Lagrangian radii for the supernova ejecta in Figure
\ref{fig:lagrangian_radii}, right panel, show first signs of
interaction of the ejecta with the circumstellar medium around
$1-10\,\textrm{kyr}$ after the explosion.  What follows is a similar
picture of expansion eventually reversing into fallback.  Fractions of
$(3.125\%,\, 6.25\%,\, 12.5\%,\, 25\%,\, 50\%)$ of the ejecta mass
turn around and start falling back at around $ (0.2,\,1.0,
\,4.0,\, 8.0,\, 30.0)\,\textrm{Myr}$ after the explosion.  The
corresponding turnaround radii are
$(3,\,8,\,30,\,60,\,200)\,\textrm{pc}$.  
The outer $\sim 50\%$ of the ejecta passes the virial radius and continues to travel outward until the
end of the simulation.  The net outflow of ejecta at $r\gtrsim
R_{200}$ can be contrasted with the net inflow of baryons at the same
radii, underscoring the three-dimensional nature of the flow and an
inadequacy of one-dimensional integrations 
in treating the long-term dynamics of supernova remnants in cosmic
minihalos.  Spherical symmetry implicit in one-dimensional
integrations forces the ejecta and the swept up circumstellar medium
into a spherical thin shell at late times. It does not allow for the
presence of the dense, low-entropy clouds that enter the halo by
infall from 
the filaments of the cosmic web. The expanding shell may pass
and leave behind such clouds. 
The spherical symmetry also does not allow radial segregation of the
ejecta as a result of the instability of the decelerating thin
shell. 

Rayleigh-Taylor fingering is evident in Figure \ref{fig:projection} at
$8.5\,\textrm{Myr}$ after the explosion.  The long term evolution of
the fragmented shell is seen in Figure
\ref{fig:metal_fallback}, left panel, which shows the ejecta tracer particles in
projection, superimposed on a slice of the density field through the
center of the halo.  It is also clear in these figures that the blastwave
has expanded biconically perpendicular to the sheet-like baryonic
overdensity deriving from the cosmic web; this pristine gas, unpolluted by
the ejecta, is still able to stream into the halo center.  In the right
panel of Figure \ref{fig:metal_fallback}, showing the central
$360\,\textrm{pc}$, pristine baryonic streams arrive diagonally
into the central few tens of parsecs, where they form a two-armed
spiral and join a
self-gravitating central
core.  At this point in the simulation, the spatial resolution is
insufficient to resolve the internal structure of the core.

Strikingly, we observe filamentary accretion of ejecta-enriched gas
perpendicular to the pristine sheet.  This is the return of the
ejecta from the supernova remnant back into the center of the halo.  A
closer examination of the time evolution of the ejecta kinematics
shows that in the snowplow phase, the ejecta are compressed into a thin
shell, as expected. The thickness of the shell is itself not 
well resolved, thus we are not in the position to explore its
susceptibility to direct gravitational fragmentation \citep[e.g.,][]{Mackey:03,Salvaterra:04,Machida:05,Chiaki:12}.
The Rayleigh-Taylor instability buckles and redistributes momentum in
the shell. The highest momentum fragments continue semi-ballistically
outward, consistent with the observation by \citet{Whalen:08b} that the
blastwave of
even a moderate $10^{51}\,\textrm{ergs}$ explosion has enough momentum to
escape a standard minihalo.  The low-momentum fragments, however, do not have
sufficient momentum to escape; they turn around, fall back, and
eventually join the central unresolved core.  The falling ejecta make
thin multiply-folded streams and filaments, indicating what seems to
be a certain lack of mixing with the 
halo gas.   The pristine gas accreting through the filaments of the
cosmic web, and the ejecta-enriched gas accreting from the supernova
remnant, meet in the central core.  If the core is turbulent---poor
resolution at this stage of the simulation would not allow us to
detect such turbulence---turbulent mixing of the ejecta, now
containing some swept
up pristine gas, with a much
larger mass of pristine gas would ensue.

\subsection{Metal Enrichment}
\label{sec:enrichment}

\begin{figure}
\begin{center}
\includegraphics[width=0.49\textwidth]{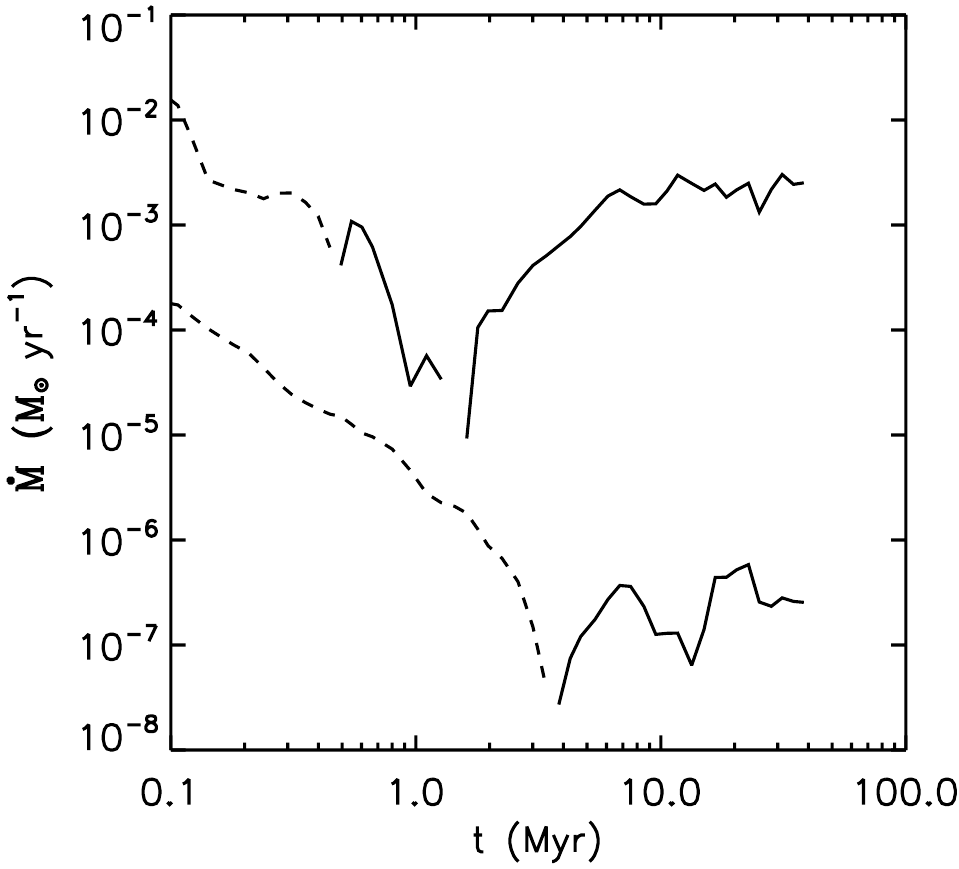}
\end{center}
\caption{Metal (lower curve; $\dot{M}_Z$) and total baryon (upper
  curve; $\dot{M}_{\rm b}$) net mass
  flow rate through a sphere of radius $20\,\textrm{pc}$ centered on
  the gravitational potential minimum.  
Dashed lines indicate outflows and solid lines inflows.  The total mass of ejecta in the
  simulation is
  $M_{\rm SN}=40\,M_\odot$ and the metal mass is $M_Z=Z_{\rm SN}M_{\rm
    SN}=4\,M_\odot$. Net outflow reverses into an inflow earlier in the
  baryons because of the presence of cold filaments delivering
  metal-free gas from the cosmic web into the halo center.\label{fig:mdot}}
\end{figure}

In Figure \ref{fig:mdot}, we show the net rates of baryon and
metal outflow from or inflow into the central $20\,\textrm{pc}$ of the
halo.  Some baryonic inflow is already evident early on, after
$50\,\textrm{kyr}$ from the explosion, and after $\approx 1.5\,\textrm{Myr}$, baryons
transition to permanent inflow.  After $\approx 5\,\textrm{Myr}$, the
inflow rate settles at $\dot M_{\rm b}\approx
0.002\,M_\odot\,\textrm{yr}^{-1}$.  The metal flow transitions from
outflow to inflow at $\approx 4\,\textrm{Myr}$ after the explosion,
and fluctuates in the range $\dot M_Z\sim
(0.5-5)\times10^{-7}\,M_\odot\,\textrm{yr}^{-1}$.  The average asymptotic
absolute metallicity of the inflowing material is 
\beq
\label{eq:metallicity_accreting}
Z_{\rm inflow}= \frac{\dot M_Z}{\dot M_{\rm b}} \sim
10^{-4}\left(\frac{Z_{\rm SN}}{0.1}\right) \sim 0.005 \,Z_\odot
\left(\frac{Z_{\rm SN}}{0.1}\right) ,
\eeq
where we are referring to the total metal mass
ignoring fractional abundances of individual metal species.

We proceed to examine the evolution of the metal-mass-weighted
metallicity probability density function (PDF).  The metallicity PDF is
computed in three different ways.  
In the first, ``fluid-fluid'' method, the metallicity is
obtained by summing the amplitudes of the passive
scalars defining metal abundances 
on the computational grid.  The metal mass at
metallicity $Z$ is then calculated by computing a volume integral of
$Z\rho$ over the cells with metallicities
in a narrow logarithmic bin containing $Z$. The shortcoming of this
method is that it yields, as a consequence of a well-known shortcoming
of passive scalar transport schemes, nonvanishing, small metallicities even in
many cells that ejecta could not have reached on hydrodynamic 
grounds.\footnote{The spurious diffusion can easily be understood by
  considering a step-function
compositional discontinuity $Z(x)=H(x)$ in an otherwise uniform medium traveling
in the $x$-direction with velocity $v_x>0$, and let $\Delta x$ be the
grid cell size. The code approximates the solution to the transport
equation $\partial Z/\partial t+v\partial Z/\partial x=0$ by computing
cell boundary fluxes with the piecewise-parabolic method; here, we adopt a simpler
scheme after assuming that just to the right of the discontinuity, the
metallicity of every cell is much smaller than that of the cell
immediately to the left. Let $\Delta t$ denote the computational
time step and $\delta\equiv v \Delta t/\Delta x$ the 
dimensionless Courant
parameter for fluid velocity, and assume that the discontinuity is slow, $\delta\ll 1$. 
Then it is straightforward to show that after $t/\Delta t$ steps,
the metal-free region to the right of the traveling discontinuity will
develop a spurious tail of nonzero metallicities of
the asymptotic form 
\beq
Z(x) \sim { t/\Delta t \choose x/\Delta x } \, \delta^{x/\Delta x} ,
\eeq
where the first factor is the binomial coefficient. The tail
establishes itself with an unphysical speed $\sim \Delta x/\Delta t \gg v$.}  With this in mind,
the first method is more accurate where the metallicity is near its
peak value, and is highly inaccurate elsewhere.  Our remaining two methods
for computing the metallicity PDF are designed to alleviate the impact
of spurious diffusion.

In the second,
``fluid-particle'' method, the metallicity $Z$ of a computational cell is calculated in the same
way as in the first method, but the metal mass is calculated by summing
the mass of the ejecta tracer particles in the cell,
and multiplying by the ejecta absolute metallicity $Z_{\rm SN}=0.1$.  The tracer
particle mass inside a cell is calculated with the cloud-in-cell method, with
the weighting function taken to correspond to spreading the particle
mass over a cubical kernel identical to the host cell, but centered on
the particle.
In the second method, only the cells having nonvanishing overlap with
at least one tracer particle cloud-in-cell kernel
contribute to the metallicity PDF.  It should be kept in mind,
however, that a metal particle displaced from the center of its host
cell contributes mass to the neighboring cell that may be separated
from the host cell by a shock transition or a contact discontinuity.

In
the third, ``particle-particle'' method, the metallicity inside the cell is itself computed
by summing up the metal mass inside the cell (again, as in the second
method, computed by adding up tracer particle mass inside the cell
and multiplying by $Z_{\rm SN}$), and dividing this by the sum of the metal
mass and the fluid hydrogen and helium masses in the cell.  Since
the combined abundance of hydrogen and helium is near unity, the
third method completely ignores the fluid metallicity as defined by
the passive scalar advected on the computational grid.  Ejecta 
particles as tracers of metallicity obviously do not suffer from the
same spurious diffusion as the fluid metallicity, but they are
affected by systematic issues of their own.  In convergent flows such
as near the insufficiently well resolved snowplow shell, metal particles can get trapped in discontinuous
flow structures, which can produce a spurious \emph{increase} of local
metallicity as computed by the third method. Thus, while the fluid
metallicity is bounded from above by its maximum value, $Z_{\rm
  SN}=0.1$, at the point of injection, the particle-based metallicity
can exceed this value and even approach unity.

\begin{figure*}
\begin{center}
\includegraphics[width=0.9\textwidth]{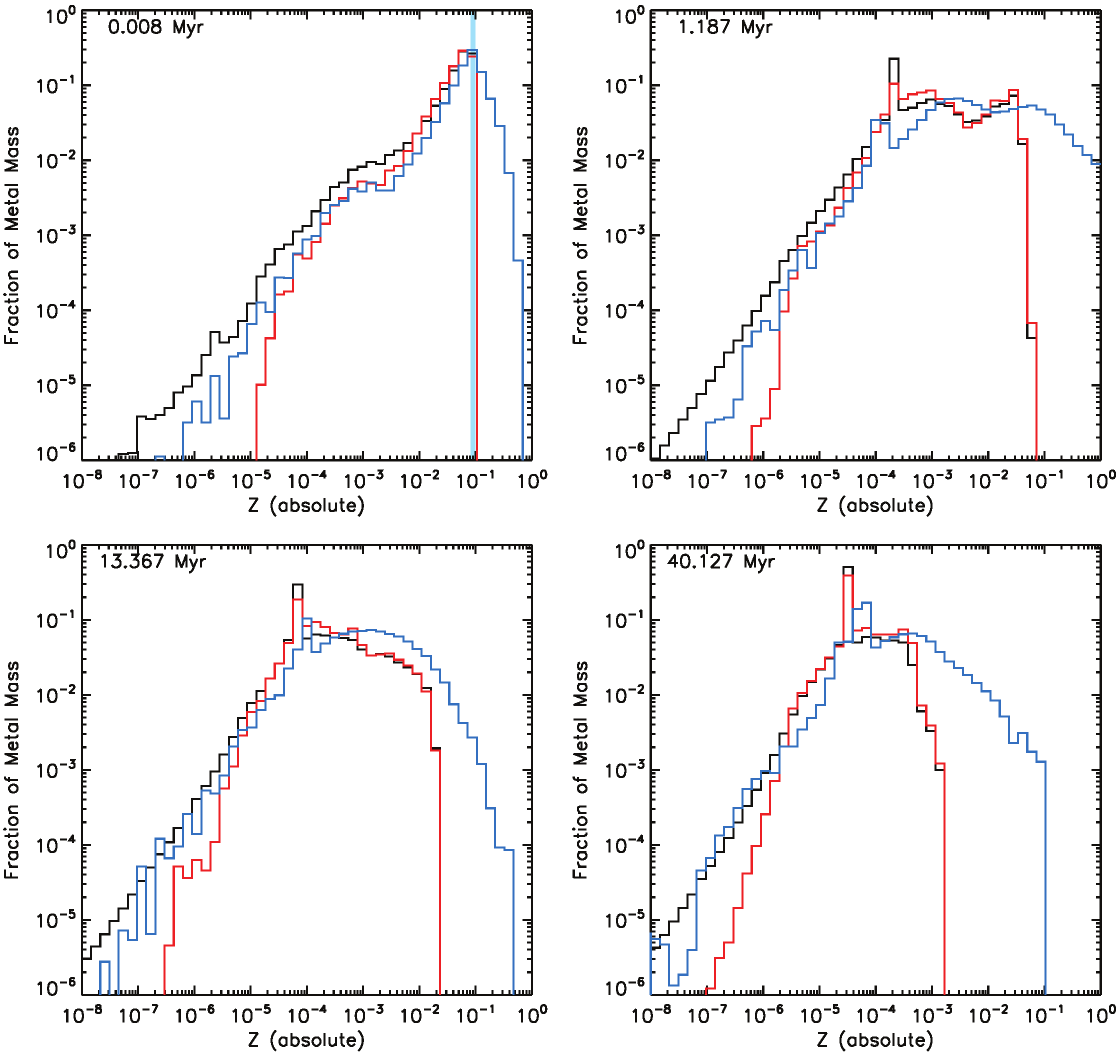}
\end{center}
\caption{Absolute metallicity distribution at four different times after the
  supernova.  The histograms show the fraction of total metal mass as
  a function of absolute metallicity calculated in three different
  ways: by computing the metallicity and metal mass from the fluid
  variables defined on the AMR grid (black curve), by computing
  metallicity from the fluid variables and metal mass from the
  Lagrangian particles tracing the supernova ejecta assuming ejecta
  metallicity of $Z_{\rm SN}=0.1$ (red curve), and by computing both
  metallicity and metal mass with the Lagrangian particles (blue
  curve).  The blue column in the first histogram shows the initial
  metallicity of the ejecta. The peak at $Z\approx (2,\,0.7,\,0.3)\times10^{-4}$ at
  $t-t_{\rm SN}=(1.187,\,13.367,\,40.127)\,\textrm{Myr}$ contains the fluid that
  has become incorporated in the central unresolved core (please see text for additional
  explanation).  \label{fig:histograms}}
\end{figure*}

In Figure \ref{fig:histograms}, we show the metal-mass-weighted
metallicity PDF at four different
times: first, in the Sedov-Taylor phase, and, then at approximately $1$, $13$, and
$40\,\textrm{Myr}$ after the explosion.  It is immediately clear that
already at early times, the metallicity PDF exhibits a tail $dM_{\rm
  Z}/d\ln Z\propto Z$ extending
toward very low metallicities, where differences between the PDFs computed
with the three methods became more severe.  Spurious diffusion
across compositional discontinuities in the fluid-based methods, and
cloud-in-cell smearing in particle-based methods, both contribute
to the tail.  We also notice that the particle-particle method has
developed a tail extending into the forbidden region, $Z>Z_{\rm SN}$,
a clear signature of spurious particle trapping.

As the supernova remnant ages, metal mixing in the complex
interior flow, which is numerical in the simulation but can be facilitated
by turbulence in nature \citep[e.g.,][]{Pan:11},
broadens the metallicity peak from its initial
location at $Z\approx Z_{\rm SN}$ to an approximate range $10^{-5}\lesssim Z
\lesssim Z_{\rm max}(t)$, where $Z_{\rm max}(t)$ is a time- and
PDF-extraction-method-dependent maximum absolute metallicity.  For fluid-based
metallicity PDFs, we find that $Z_{\rm max}\approx
(0.07,\,0.02,\,0.002)$ at $t-t_{\rm SN}\approx
(1,\,13,\,40)\,\textrm{Myr}$. We also notice a prominent narrow
metallicity peak, built   from cells belonging to the central,
unresolved core that is accreting both from the pristine filaments of
the cosmic web and from the ejecta-enriched supernova remnant
fallback.  The peak is migrating slowly toward lower metallicities, covering
$Z_{\rm core}\approx (2,\,0.7,\,0.3)\times10^{-4}$ in the same three epochs. The
  metallicity of the unresolved core is consistent with the average
  metallicity of the accreting fluid given in Equation (\ref{eq:metallicity_accreting}).

\begin{figure}
\begin{center}
\includegraphics[width=0.49\textwidth]{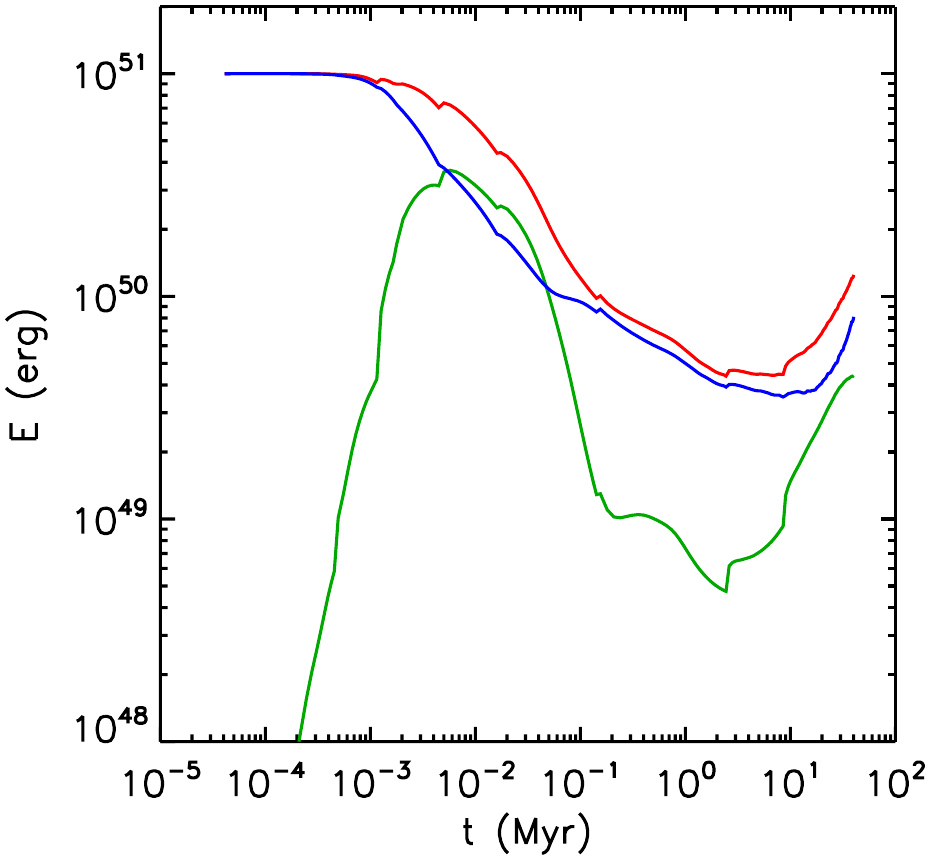}
\end{center}
\caption{The kinetic (blue curve), internal (green curve), and combined
  kinetic and internal (red curve) energy of the metal enriched fluid, defined
  as the fluid with absolute metallicity $Z\geq 10^{-6}$. \label{fig:energy}}
\end{figure}

Finally, we turn to the energetics of the metal-enriched material.  In
Figure \ref{fig:energy}, we show the evolution of the total kinetic and
internal (thermal) energy of fluid elements having absolute
metallicity $Z\geq 10^{-6}$.  At $1\,\textrm{kyr}$ after the
explosion, 
the kinetic energy
 decreases below its initial value when the ejecta start interacting with the ambient
  medium and the well-resolved reverse shock facilitates kinetic-to-internal
  conversion. Adiabatic and radiative cooling both contribute to the
  rapid decrease of internal energy at $0.01-1.0\,\textrm{\rm Myr}$
  after the explosion.  The renewed rise of both energies
  in the old remnant reflect the fallback of ejecta-enriched fluid into
  the halo center. 

\section{Discussion}
\label{sec:discussion}

Existing treatments of metal injection by Pop III supernovae have
followed the 
excursion of the metal-endowed ejecta into the cosmic
environment and their possible fallback into dark matter halos, but
usually with one-dimensional simulations
\citep{Kitayama:05,Whalen:08b} or relatively low-resolution 
three-dimensional, grid-based or smoothed particle hydrodynamic
simulations \citep{Nakasato:00,Bromm:03b,Greif:07,Wise:08b}.  These treatments
point to a prompt and widespread dispersal of supernova
ejecta into the primordial IGM if one assumes that the
supernovae were extraordinarily energetic, e.g., consistent with the popular
hypothesis that the Pop III stars exploded as PISNe.
However, in particularly
high mass halos or those hosting moderate mass stars,
the detailed dynamics of the ejecta is sensitive to the
pre-supernova evolution of the \ion{H}{2} region.
With one-dimensional hydrodynamic simulations,
\citet{Whalen:08b} showed that the \ion{H}{2} region of a
moderate mass star at the center of the minihalo may remain confined
within the minihalo.  They calculated that the blastwave of a $10^{51}\,\textrm{ergs}$
supernova in $\lesssim 10^6\,M_\odot$ minihalos, does, after $\sim 10\,\textrm{Myr}$,
reach the virial radius of the halo, and concluded that this results in
the ``disruption'' of the baryonic component of the halo. They also
considered high-mass $\sim 10^7\,M_\odot$ minihalos in which a strong
photodissociating background had prevented star formation
\citep[e.g.,][]{OShea:08}, and found that there the \ion{H}{2} region
may not break out and the supernova blastwave may be
confined at very small radii.  They speculated that the subsequent
prompt fallback of metal-enriched supernova ejecta into the center of the
halo could lead to the formation of a globular-cluster-like stellar
system, similar to the suggestion of \citet{Wada:03} that a relatively
small
number of supernovae exploding in a more massive, $\sim10^8\,M_\odot$
cosmic halo could also lead to a fallback and prompt star
cluster formation.  This is a particularly interesting possibility, in view of the
unexplained origin of metal-poor globular clusters \citep{Brodie:06}: do they form in
the centers of their own low-mass dark matter halos \citep[e.g.,][]{Peebles:84,Rosenblatt:88,BrCl:02,Boley:09,Griffen:10}, or do
they form
off-center in the merging and fragmentation of more massive
protogalactic disks \citep[e.g.,][and references
therein]{Kravtsov:05}?  

It is important to
realize that the fate of the supernova ejecta may be
influenced by dynamical processes operating on relatively small
spatial scales and strongly violating spherical symmetry.  In addition
to the likely intrinsic asphericity of the explosion itself, one
such process is the Rayleigh-Taylor instability of the decelerating contact
discontinuity after the ejecta have started to interact with the
ambient medium. The ambient medium into which the supernova expands is
bound to be inhomogeneous, especially if the luminosity of the
progenitor stars and other hot stars that have formed with it is
insufficient to completely photoionize the densest clumps remaining
from the progenitor gas cloud.  On
the somewhat larger $\sim100\,\textrm{pc}$ scales of the host
minihalo, a pronounced departure from spherical symmetry will arise from
the filamentary geometry of the cosmic web. Such
asymmetry can interact with the expanding blastwave.  Further, as an aging
supernova remnant forms a thin, momentum-conserving snowplow shell, the
decelerating shell is itself unstable to the Rayleigh-Taylor
instability and forms fingers.  Finally, while the blastwave
instabilities lead to partial entrainment of the ejecta and the shocked
primordial gas, any ejecta-enriched gas falling back toward the center
of the halo, where it may engage in a new round of star formation, may
not necessarily be chemically homogeneous (full chemical mixing requires at
least a few turbulent eddy turnover times which may not be available in
the blastwave).  Most of these phenomena remain to be explored in the
context of Pop III stars; the results we have presented here are one
of the first steps in that direction.

%The simulation shares some similarities with that of \citet{Boley:09}
%and its setup has many common elements with that of  \citet{Wise:12}. The latter work followed the evolution of
%a cosmological region of the same size as ours but to much lower
%redshifts, with a focus on the statistics of star formation and metal
%enrichment across a large number of star-forming halos.  Our focus,
%instead, is on the detailed, smaller-scale character of metal
%transport in the aftermath of a single supernova.

\subsection{Metal-Enriched Star Formation: Continuous, Bursty, or Self-Limiting?}

The prompt resumption of baryonic infall into the
center of the cosmic minihalo suggests that star formation 
could resume on similarly short, $\sim 1-5\,\textrm{Myr}$, time scales.  In the simulation
presented here, the metals that
the supernova has synthesized begin returning to the halo center
$\sim4\,\textrm{Myr}$ after the explosion. The mean
metallicity of the inflow, averaged over the 
primordial streams and the metal-enriched collapsing supernova remnant streams
converging
in the center of the halo,
is $\sim 0.005\,Z_\odot$ assuming a net supernova metal yield of $4\,M_\odot$.  If the
metal-rich gas successfully mixes with the primordial gas, then,
according to a prevailing belief, 
the metallicity of the combined streams is well sufficient to ensure that new
star formation in this gas should be producing normal, low-metallicity
Pop II stars. The Pop III to II transition cannot be
ascertained in the present simulation because we do
not adequately resolve the cold gas at densities $n_{\rm H}\gtrsim
100\,\textrm{cm}^{-3}$.  This gas belongs to the central core and is
colder than primordial gas at the same densities preceding the
insertion of the first star. It is at unresolved densities that the
metallicity of the gas strongly affects the thermodynamic
evolution and thus governs the outcome of fragmentation
\citep[e.g.,][assuming low metallicities, 
$Z\lesssim 0.01\,Z_\odot$]{Omukai:05,Santoro:06,Schneider:06,Smith:09,SafranekShrader:10}.

The gaseous core in the center of the minihalo that is fed by metal-enriched streams
can be compared to the star-forming core of a
giant molecular cloud (GMC).  We attempt such comparison with an eye
on the overall hydrodynamics of the gas flow, putting aside any differences in chemical
composition, dust content, CMB temperature, etc., in the two systems.   Dense GMC cores grow by accreting from 
larger-scale, lower-density cloud environments which are supersonically
turbulent.  Smaller cores and
their embedded proto star clusters merge with each other to form larger
associations.  Star formation in the cores
is thought to be self-regulated, with the ``feedback'' from ongoing star
formation controlling the amount of star-forming
gas available and eventually leading to GMC dissolution
\citep[e.g.,][]{Murray:10,Murray:11,Goldbaum:11,Lopez:11}. The
feedback entails, e.g., the pressure of the \ion{H}{2} 
regions and the pressure of
direct stellar radiation. While the baryonic mass of a cosmic minihalo is perhaps smaller than that of a
GMC, the centrally concentrated underlying dark matter
halo renders the gravitational potential depth in a minihalo's central
gaseous core similar to that in a
GMC's star-forming core.  The underlying dark matter gravitational
potential may make a cosmic minihalo less susceptible to star-formation-induced dissolution
than a GMC, assuming that the two systems are forming stars at the same rate.

GMCs are
thought to sustain star formation over tens of millions of years, in
spite of the \ion{H}{2} regions, supernovae, and other feedback activity that commences
quickly following the formation of the first massive stars.  The
feedback's impact preceding the eventual dissolution of the cloud
complex
may be limited to reducing the average star
formation rate. We do not
know whether 
star formation in metal-enriched minihalos and their descendent $\gtrsim
10^7\,M_\odot$ halos can be sustained as in GMCs in spite of the
feedback, or is the feedback from second-generation stars so 
effective as to evacuate baryons from the halo and prevent further star
formation on cosmological time scales \citep[see, e.g.,][]{Mori:02,Frebel:12}. It seems that these two
scenarios can be evaluated only with attention to the detailed,
small-scale mechanics of star formation and its feedback.   A
corollary of the first scenario would be the presence of multiple,
chemically distinct
stellar populations in the stellar system that forms this way.

The statistics of low-luminosity stellar systems
in the Local Group in conjunction with theoretical modeling of
structure formation in a $\Lambda$CDM universe
suggests that star formation must have, to some
extent, been suppressed in low-mass cosmic halos \citep[e.g.,][]{Susa:04a,Ricotti:05,Gnedin:06,Moore:06,Madau:08,Bovill:09,Bovill:11,Koposov:09,Munoz:09,Salvadori:09,Busha:10,Gao:10,Griffen:10,Maccio:10,Okamoto:10,Font:11,Lunnan:11,Rashkov:11}. This suppression, however,
could have arisen both internally, as in GMCs,
and externally, where the suppression is driven by the heating of the
IGM by the background
ionizing radiation coming from other, more massive but less common halos 
\citep[e.g.,][]{Thoul:96,Kepner:97,Barkana:99,Shapiro:04,Susa:04b,Mesinger:06,Mesinger:08,Okamoto:08}.  The
heating could have shut off the inflow of fresh gas into the halo,
eventually stunting star formation.  The prevalence of either
suppression mechanism 
as a function of halo properties cannot be
understood without first understanding the detailed
mechanics of star formation and feedback following the initial
fallback of Pop III supernova ejecta.

\subsection{Uncertainties Related to the Nature of First Stars}

A source of uncertainty regarding the
formation and nature of the first metal-enriched star clusters is the
incomplete understanding of the precise character of metal-free
Pop III stars, and of the energetics and
nucleosynthetic output of the associated supernovae.  Theoretical
investigations of Pop III star formation have not yet
converged and the precise functional form of the resulting initial mass function is not yet known.  The
evolution of metal-free stars is
very sensitive to their poorly-understood internal rotational structure
starting with the
protostellar phase \citep[see, e.g.,][]{Stacy:11,Stacy:12b}, and on the
potential presence of a close companion.
Similarly, the explosion mechanism in metal-poor stars is not known.
Each of the suggested mechanisms, including the standard
delayed-neutrino mechanism hypothesized to drive Type II explosions,
and the pair-instability and pulsational
pair-instability mechanisms expected in relatively massive metal-free
stars \citep[see, e.g.,][and references therein]{Heger:02,Chatzopoulos:12,Yoon:12}, 
as well as the more exotic ``hypernova''-type mechanisms involving
a (possibly magnetized) outflow from a central compact object, implies
its own characteristic explosion energy and nucleosynthetic
imprint.  

Another aspect of the final stellar demise that should be very
sensitive to both the stellar mass and rotation is the nature of the
compact remnant, if any, that it leaves behind.  While the pair
instability disperses the star completely, leaving no remnant, core
collapse in stars with initial masses $\gtrsim 15-25\,M_\odot$
\citep[the precise threshold mass depending on unknown aspects of the
explosion; see, e.g.,][]{Heger:03,Zhang:08} leaves behind a black hole, either
via direct collapse into one, or by fallback of the ejecta onto the
neutron star. The black hole may be produced with an initial kick that
would eject it from the dark matter halo \citep{Whalen:12}, but if the
kick is small and the black hole is retained near the
gravitational center, it could accrete the gas collecting in
this region. The energy liberated in this accretion could be another,
potentially significant source of feedback.  The accretion rate and the
luminosity of this source would be subject to complex 
radiation-hydrodynamic couplings that are only
understood under the most idealized assumptions, such as in an absence of
shadowing by or radiation trapping within the innermost accretion flow
\citep[see,
e.g.,][]{Milosavljevic:09a,Milosavljevic:09b,Park:11,Park:12,Li:11}.
To date, only very few cosmological simulations investigating the
impact of the radiation emitted by accreting first-star remnants on
hydrodynamics in the host halo have been carried out
\citep[e.g.,][]{Kuhlen:05,Alvarez:09,Johnson:11,Jeon:12} and only a partial picture of the
feedback on second generation star formation is available.  
\citet{Jeon:12} show that X-rays from the remnant
impact the cosmic neighborhood in complex ways but do not entirely
prevent (and can actually promote) gas cooling and star formation.
Much remains to be understood about the role of compact remnants in
the evolution of the first galaxies.

Many investigations of the formation of first
stars and galaxies and their role in reionization have explored implications of
the hypothesis that Pop III stars had high masses, explosion energies,
and metal yields. We find that the opposite limit of moderate
masses, explosion energies, and metal yields might imply sharply
divergent outcomes, 
not only for the parent cosmic minihalos, but also
for the overall pace of transformation of the early universe. A related issue
is the nature of the first galaxies \citep[for a review, see][]{BrYo:11}.
Their assembly process is influenced 
by the feedback from the Pop III
stars that formed in the minihalo progenitor systems, and this feedback
in turn sensitively depends on the mass scale of the first stars.
In the case of extremely massive stars, negative feedback is very disruptive,
completely sterilizing the minihalos after only one episode of star
formation. Only after $\sim 10^8\,\textrm{yr}$ did the hot gas sufficiently cool
to enable the recollapse into a much more massive dark matter halo, where
a second-generation starburst would be triggered \citep{Wise:08b,Greif:10}. This separation into
two distinct stages of star formation would be smoothed out in the
case studied here. Indeed, the first galaxies may then already have
commenced their assembly in massive minihalos, as opposed to the atomic cooling halos \citep[e.g.,][]{OhH:02}
implicated in the previous scenario.

\section{Conclusions}
\label{sec:conclusions}

We have carried out a cosmological hydrodynamical simulation designed
to investigate the evolution of a cosmic minihalo in the aftermath of
the formation of the first, Pop III star, assuming that the
star has a moderate mass and explodes as a moderate-energy Type II-like
supernova, consistent with the recent downward revision of Pop
III stellar mass estimates \citep{Stacy:10,Stacy:12a,Clark:11,Greif:11,Greif:12,Hosokawa:11}.
We analyzed the dynamics of supernova ejecta and gas flow
inside the minihalo.  These can be compared to similar studies
involving massive Pop III stars exploding as ultra-energetic
PISNe.  Our main conclusions are as follows.

The moderate-mass star and the moderate-energy supernova it produces
inflict significantly smaller damage to the host minihalo than that
inflicted by a massive star exploding as an ultra-energetic supernova.
The star only partially photoionizes the host
halo. The densest gas flowing from the filaments of the cosmic web
into the star's parent cloud remains neutral and survives
the supernova blastwave. This dense gas resumes accretion into the
center of the minihalo only $\sim1\,\textrm{Myr}$ after the
explosion.  After $\sim 20\,\textrm{Myr}$, all spherically-averaged
baryonic 
mass coordinates are moving inward and the accretion rate at
the center of the halo is $\sim 0.002\,M_\odot\,\textrm{yr}^{-1}$.

Following instability of the blastwave during the
snowplow phase, a fraction of the supernova ejecta starts falling back to the center of the
halo. The ejecta fallback accretion rate reaches a steady state after
$\sim 5\,\textrm{Myr}$.  The accreting ejecta are incompletely mixed with the
primordial gas and are confined in thin sheets and filaments.  
Less than a half of the ejecta escape the virial radius; the escaping
ejecta can be traced back to the outermost Rayleigh-Taylor fingers.

The average metallicity of the accreting matter is in the range
$0.001-0.01\,Z_\odot$. This result depends on the assumed supernova
energy and metal yield, but in this study, we did not pursue parameter
space exploration.  The metallicity of the accreted gas 
is sufficient to ensure that new stars forming in the central core,
unresolved in the simulation, will have even
lower characteristic masses.  

These results bring into focus metal-enriched star formation in cosmic minihalos and
their immediate descendent halos.  The character of post-supernova
metal transport and gravitational fragmentation of the
metal-enriched gas, and of the impact of subsequent star formation and
supernovae, are both worth addressing in follow-up investigations.
The early universe likely exhibited a range of different explosion settings,
from modest to ultra-energetic. It will be intriguing to see how these
high-redshift systems are matched with possible local fossils, such as
the ultra-faint dwarf galaxies
\citep[e.g.,][]{Bovill:09,Salvadori:09,Tumlinson:10} and metal-poor
globular clusters \citep[e.g.,][]{Brodie:06}. From what we have
already learned, Pop III supernova feedback 
appears to play a key role in shaping early cosmic history and the
nature of relic stellar systems.

\acknowledgements

We gratefully acknowledge stimulating discussions with  J.\
Scalo and helpful technical consultation with A.\ Dubey, C.\ Daley,
T.\ Plewa, P.\ Ricker, and M.\ Ruszkowski.  We thank R.\ Banerjee for
providing part of the software used to visualize this work.  J.\ S.\ R.\
was supported in part by the John W. Cox Endowment for the
Advanced Studies in Astronomy.
This research was supported by NSF grants AST-0708795 and AST-1009928
and NASA ATFP grant NNX09AJ33G.  The authors acknowledge the Texas
Advanced Computing Center (TACC) at The University of Texas at Austin
for providing computing and visualization resources that have contributed to the research results reported within this paper. The software used in this work was in part developed by the DOE NNSA-ASC OASCR Flash Center at the University of Chicago.

\end{document}